\documentclass[twocolumn,showpacs,prl,nofootinbib,superscriptaddress,preprintnumbers,amsmath,amssymb]{revtex4}

\usepackage{graphicx,subfigure}
\usepackage{dcolumn}
\usepackage{bm}
\usepackage{color}
\usepackage{rotating}
\usepackage{ulem} 
\usepackage{float}

\newcommand{\galprop}{{\sc GALPROP}}
\newcommand{\GP}{{\galprop}}

\newcommand{\frankie}{{\sc FRaNKIE}}

\newcommand{\gray}{$\gamma$-ray}
\newcommand{\grays}{$\gamma$-rays}

\begin{document}

\preprint{}

\title{Galactic PeVatrons and helping to find them: Effects of Galactic absorption on the observed spectra of very high energy \gray{} sources}

\author{T.~A.~Porter}
\affiliation{W. W. Hansen Experimental Physics Laboratory and Kavli Institute for Particle Astrophysics and Cosmology, Stanford University, Stanford, CA 94305, USA}
\author{G.~P.~Rowell}
\affiliation{School of Physical Sciences, University of Adelaide, Adelaide, SA 5000, Australia}
\author{G.~J\'ohannesson}
\affiliation{Science Institute, University of Iceland, IS-107 Reykjavik, Iceland and Nordita, KTH Royal Institute of Technology and Stockholm University
Roslagstullsbacken 23, SE-106 91 Stockholm, Sweden}
\author{I.~V.~Moskalenko}
\affiliation{W. W. Hansen Experimental Physics Laboratory and Kavli Institute for Particle Astrophysics and Cosmology, Stanford University, Stanford, CA 94305, USA}
\date{\today}

\begin{abstract}
  Identification of the cosmic-ray (CR) ``PeVatrons'', which are sources capable of accelerating particles to $\sim10^{15}$~eV energies and higher, may lead to resolving the long-standing question of the origin of the spectral feature in the all-particle CR spectrum known as the ``knee''.
  Because CRs with these energies are deflected by interstellar magnetic fields identification of individual sources and determination of their spectral characteristics is more likely via very high energy \gray{} emissions, which provide the necessary directional information.
  However, pair production on the interstellar radiation field (ISRF) and cosmic microwave background (CMB) leads to steepening of the high-energy tails of \gray{} spectra, and should be corrected for to enable true properties of the spectrum at source to be recovered.
  Employing recently developed three-dimensional ISRF models this paper quantifies the pair-absorption effect on spectra for sources in the Galactic centre (GC) direction at 8.5~kpc and 23.5~kpc distance, with the latter corresponding to the far side of the Galactic stellar disc where it is expected that discrimination of spectral features $>10$~TeV will be possible by the forthcoming Cherenkov Telescope Array (CTA).
  The estimates made suggest spectral cutoffs could be underestimated by factors of a few in the energy range so far sampled by TeV \gray{} telescopes.
  As an example to illustrate this, the recent HESS measurements of diffuse \gray{} emissions possibly associated with injection of CRs nearby Sgr~A$^*$ are ISRF-corrected, and estimates of the spectral cutoff are re-evaluated.
  It is found that it could be higher by up to a factor $\sim 2$, indicating that these emissions may be consistent with a CR accelerator with a spectral cutoff of at least 1\,PeV at the 95\% confidence level.
\end{abstract}

\maketitle

{\it Introduction:} The origin of the spectral feature known as the ``knee'' in the CR spectrum around $\sim 10^{15}-10^{16}$ eV is not fully resolved yet.
It is thought to be a signature of the transition from predominantly Galactic to predominantly extragalactic CRs \citep[e.g.,][]{1964ocr..book.....G,1990acr..book.....B}.
Calculations employing self-consistent magnetohydrodynamic models indicate that amplification of the magnetic field by the streaming instability generated by escaping CRs ahead of supernova remnant shocks (SNRs) may facilitate particle acceleration by young SNRs into the multi-PeV region \citep[][and references therein]{2001MNRAS.321..433B,2013MNRAS.431..415B}.
Indeed, measurements of the spectra of individual elements below the knee show that they are well aligned, while their cutoff energy depends on the nucleus charge thus supporting the hypothesis of the Galactic origin of CRs up to $\sim$$10^{17}$~eV \citep[][and references therein]{2016A&A...595A..33T}.
Therefore, the knee has a complex structure where the elemental abundances may change dramatically over a relatively narrow energy range.

Identifying unique sources of these particles within the Galaxy, the so-called CR ``PeVatrons'', is effectively impossible with CR data alone because the scattering off magnetic turbulence in the interstellar medium (ISM) and escape from the Galaxy alters the initial spectrum and scrambles directionality.
Instead secondary messengers like \gray{s}, which are produced by interactions of the CRs with gas and radiation nearby their source regions, are undeflected by magnetic fields and trace directly to their origin near the CR sources.

The detection by the HESS instrument of \gray{} emissions extending well beyond 10~TeV about the GC \citep{2016Natur.531..476H} has been suggested as the first direct evidence for an individual source of CRs with energies $\gtrsim 1$~PeV.
  The measured profile indicates that it is likely due to continuous injection of CR protons over the last $\sim 10^4$ years associated possibly with the central black hole Sgr~A$^*$, or other nearby particle injector~\citep{2016Natur.531..476H,2018arXiv180402331A}.
  The alternative leptonic-induced explanation has difficulty in matching the
  hard \gray{} flux $\gtrsim 10$~TeV, meanwhile determination of the spectrum nearby the source may provide further clues to its origin.
The \gray{} spectra as detected at Earth include both intrinsic and extrinsic effects: the CR acceleration and local conditions shape the spectrum in and about the injection region \citep[e.g.,][]{2017ApJ...836..233G}, while the absorption of \gray{s} in the ISM via pair production on the ISRF and CMB provide further spectral modification.
For the CR proton injection scenario for the HESS GC source, the maximum particle energies from fitting to data span $\sim400$~TeV (95\% confidence) to $\sim3$~PeV (68\% confidence), but the effect of pair absorption on these cutoff energy estimates is not evaluated by \citet{2016Natur.531..476H}.

Attenuation on the CMB provides a $\sim$~kpc-scale \gray{} ``horizon'' for energies $\gtrsim 1000$~TeV \citep[e.g.,][]{1986MNRAS.221..769P}.
In the range $\sim 300-1000$~TeV spectral softening due to absorption on the CMB is solely dependent on the distance of a source from the Earth.
For energies below this the Galactic ISRF is the absorber with the majority effect occurring in the energy range $\sim 10-200$~TeV with its spatial dependence first calculated by \citet{2006ApJ...640L.155M} and \citet{2006A&A...449..641Z} using the 2D Galactocentric symmetric ISRF model of \citet{2005ICRC....4...77P}.
The work of \citet{2006ApJ...640L.155M} also included the effects associated with anisotropic angular distribution of the ISRF at each point in the Galaxy.
Recent re-evaluations of the pair-absorption effect using alternative 2D-based ISRF models \citep{2016PhRvD..94f3009V,2017MNRAS.470.2539P} find comparable levels to the earlier calculations.
Correction of observed \gray{} spectra for the absorption is necessary to recover intrinsic spectral characteristics for the VHE \gray{} sources.

In this paper, two recently developed 3D models of the Galactic ISRF \citep{2017ApJ...846...67P} are used to calculate the pair-absorption optical depth using the full angular distribution of the background photons for each model.
Each ISRF model is evaluated using the same radiation transport code but they use different stellar luminosity and dust density distributions.
However, their predicted local intensities are consistent with near- to far-infrared observations, and both represent current state-of-the-art solutions for the Galaxy-wide low-energy photon spectral intensity distribution.
Due to the different stellar/dust distributions for each ISRF solution the predicted photon densities elsewhere than the local region, particularly over the inner Galaxy, are not the same. 
The F98 model (see below for the meaning of their designations) gives an estimate for the strongest infrared emissions over this region, while the R12 model provides close to a lower bound (why is discussed further below).
Because it is the infrared photon density that predominantly determines the attenuation by pair creation at 10--100~TeV energies, the models provide bounds on its effect on \gray{} spectra.

The optical depths calculated for each ISRF model are used to correct the HESS observations of \gray{s} that may be produced by CRs injected nearby Sgr~A$^*$ and diffusing about the GC.
The ISRF-corrected data are then refit to obtain new estimates for the intrinsic cutoff energy for the CR proton injection scenario considered as a likely origin for the \gray{} emissions reported by \citet{2016Natur.531..476H}.
While it is not certain if Sgr~A$^*$ is actually responsible for the injection of these particles, the ISRF-correction applies independent of the scenario, and the source specifics are not important in terms of the change in maximum particle energies.

{\it Calculations:} The pair absorption optical depth for \grays\ is given by the general formula
\begin{equation}
\tau_{\gamma\gamma}(E)= \int_L dx \int d\varepsilon \int d\Omega \,
\frac{dJ(\varepsilon,\Omega,x)}{d\varepsilon d\Omega} \, \sigma_{\gamma\gamma}
(\varepsilon_c) (1-\cos\theta),
\label{eq:tau}
\end{equation}

\noindent
where $dJ(\varepsilon,\Omega,x)/d\varepsilon d\Omega$ is the 
differential intensity of background photons at the 
point $x$, $\varepsilon$ is the background photon energy, 
$d\Omega$ 
is a solid angle, $\sigma_{\gamma\gamma}$ is the Klein-Nishina cross section for the process $\gamma\gamma\to e^+e^-$ \citep{1976tper.book.....J},
$\varepsilon_c=[\frac12\varepsilon E (1-\cos\theta)]^{1/2}$
is the centre-of-momentum system energy of a photon, 
and $\theta$ is the angle between the 
momenta of the two photons in the observer's frame. 
The integral over $x$ should be taken along the path of the \grays\ 
from the source to the observer.
The calculations explicitly take into account the angular distribution of the ISRF photons, which produces noticeable effects compared to the usually employed isotropic approximation \citep{2006ApJ...640L.155M}.

Equation~(\ref{eq:tau}) is evaluated over the sky using HEALPix Nside = 64 maps for \gray{} energies from 1~to~1000~TeV with 4 logarithmic bins/decade for distance bins 0.5~kpc spacing outward from the Solar system. 
The background photon distribution $J(\varepsilon,\Omega,x)$ has two components: the ISRF and the CMB.

\begin{figure*}[ht!]
  \subfigure{
    \includegraphics[scale=0.85]{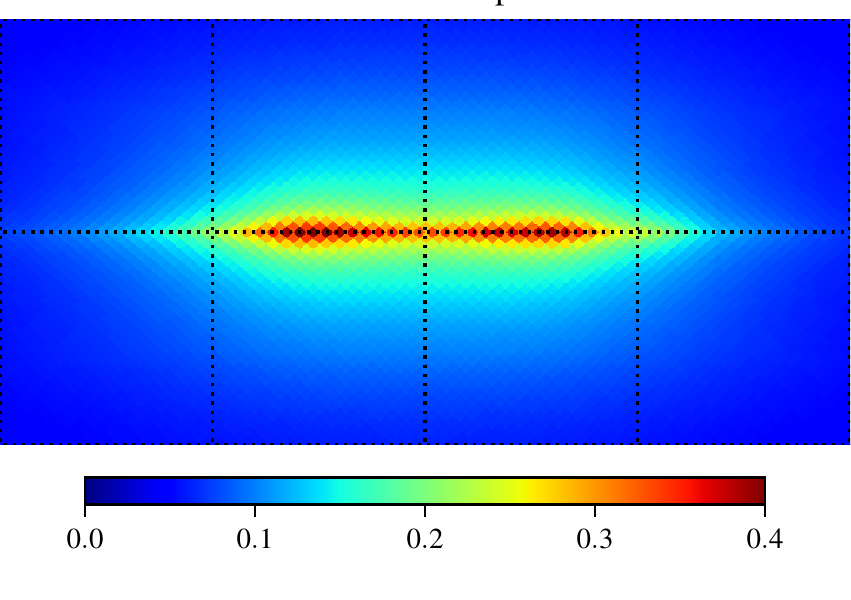}
    \includegraphics[scale=0.85]{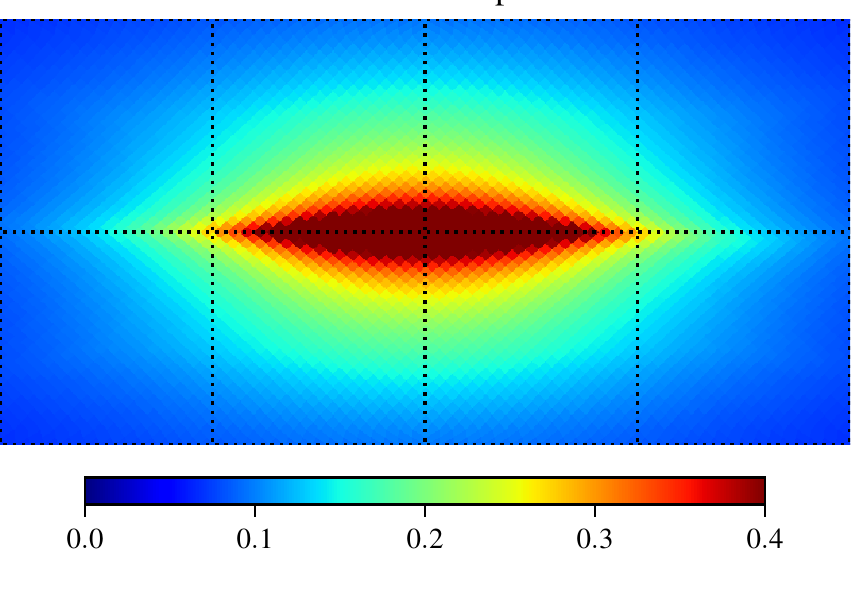}
  }
  \caption{Optical depth (Eq.~\ref{eq:tau}) at 100~TeV for the R12 (left) and F98 (right) ISRF models for an integration path length of 23.5~kpc. 
    The maps are in cartesian projection covering Galactic coordinates $-90^\circ\leq l\leq 90^\circ$ and $-45^\circ\leq b\leq 45^\circ$ with $l,b = 0^\circ, 0^\circ$ at the centre.
The longitude meridians and latitude parallels have $45^\circ$ spacing. Note that the scale saturates at an optical depth of 0.4 for ease of visual comparison between the R12 and F98 predictions (the F98 model produces optical depths higher than 0.4 for this distance horizon toward the inner Galaxy).
  \label{fig:tau}}
\end{figure*}

The spectral intensities for the ISRF are taken from \citet{2017ApJ...846...67P}, who used the Fast Radiation transfer Numerical Kalculation for Interstellar Emission (\frankie) code to calculate two Galaxy-wide distributions
based on different 3D stellar and dust density models.
The ISRF models are designated R12 and F98 following the spatial densities of stars and dust used for the radiation transfer calculations: the R12 model is based on the work of \citet{2012A&A...545A..39R} and includes spiral arms and a `hole' in the dust distribution for Galactocentric radius $R \lesssim$~3.5~kpc, while the F98 model is based on the analysis by \citet{1998ApJ...492..495F} and has an asymmetric stellar bar and a smaller `hole' in the dust distribution.
The parameters of the models are adjusted so that the local near- to far-infrared data are reproduced (see Sec.~3.2 of \citep{2017ApJ...846...67P} for the detailed description and comparison of models with data).
Because of the strong effect of dust reprocessing on the transmitted stellar light there is some degeneracy between the stellar luminosity and dust density distributions that produces variance in the modelled photon density distributions, particularly over the inner Galaxy.
For example, because of its very low dust density over the inner Galaxy the R12 model predicts correspondingly weak infrared emissions from this region, and the bulk are from where the stellar spiral arms and dust density distributions peak for $R\sim4-5$~kpc.
Meanwhile, the F98 model has a much higher dust density over the inner Galaxy and consequently produces more infrared emissions; the difference between the modelled photon densities at far-infrared wavelengths is about an order of magnitude.
But the components of either model cannot be arbitrarily modified because they are already providing a reasonable agreement with the data from near to far-infrared wavelengths.
So, at the present stage of modelling, the R12 and F98 models can be considered to provide reasonable bounds on the intensity distributions of low-energy photons in the Galaxy.
For VHE \gray{} sources located toward, or beyond, the inner Galaxy the variance in the predicted infrared photon distributions will produce correspondingly different pair-absorption effects motivating the consideration of both models in this paper.

\begin{figure*}[ht!]
  \subfigure{
    \includegraphics[scale=0.95]{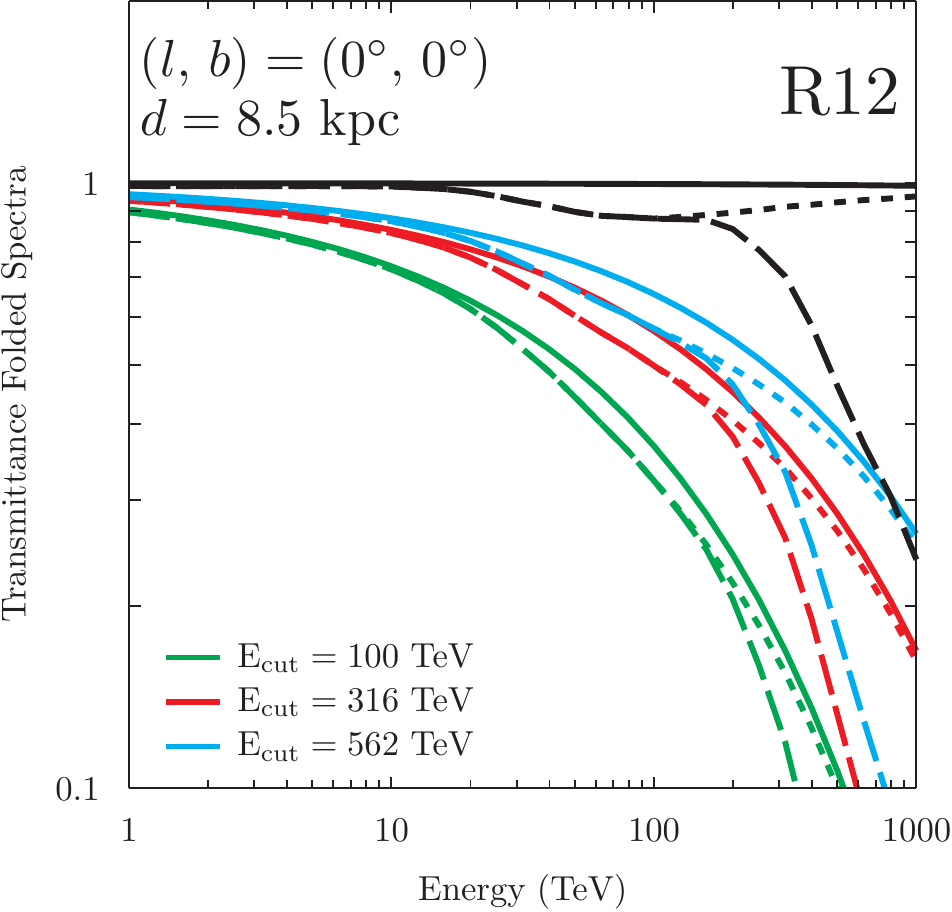}
    \includegraphics[scale=0.95]{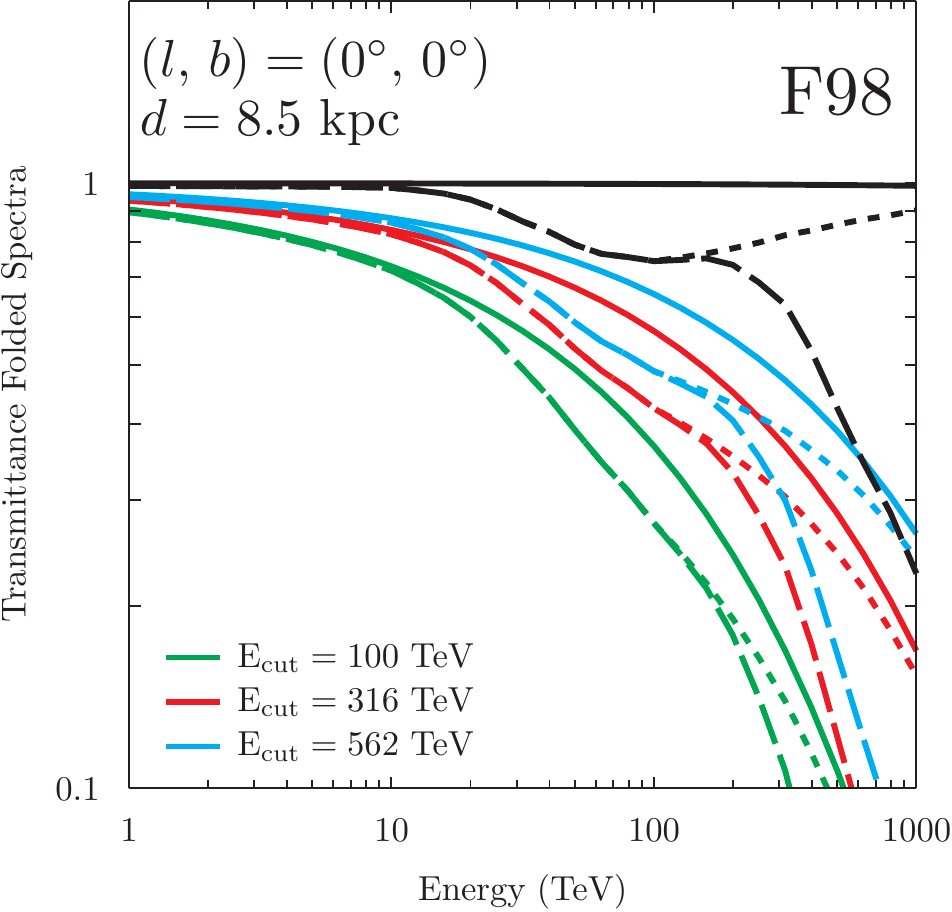}
  }
  \subfigure{
    \includegraphics[scale=0.95]{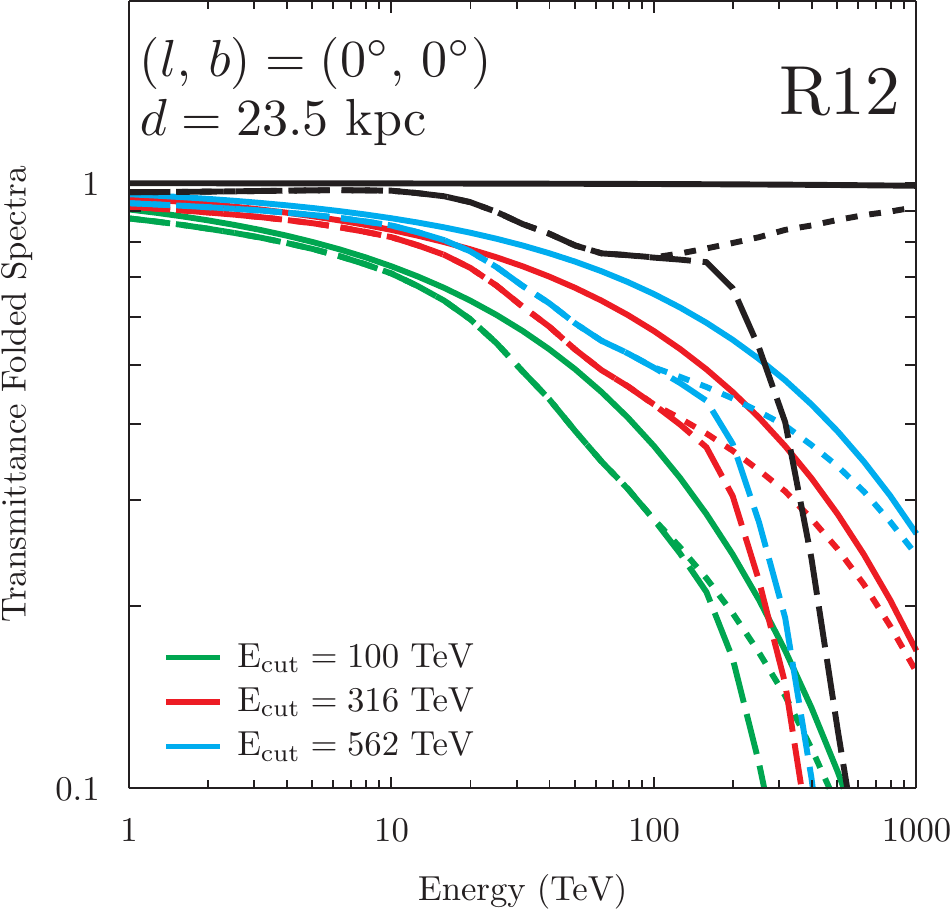}
    \includegraphics[scale=0.95]{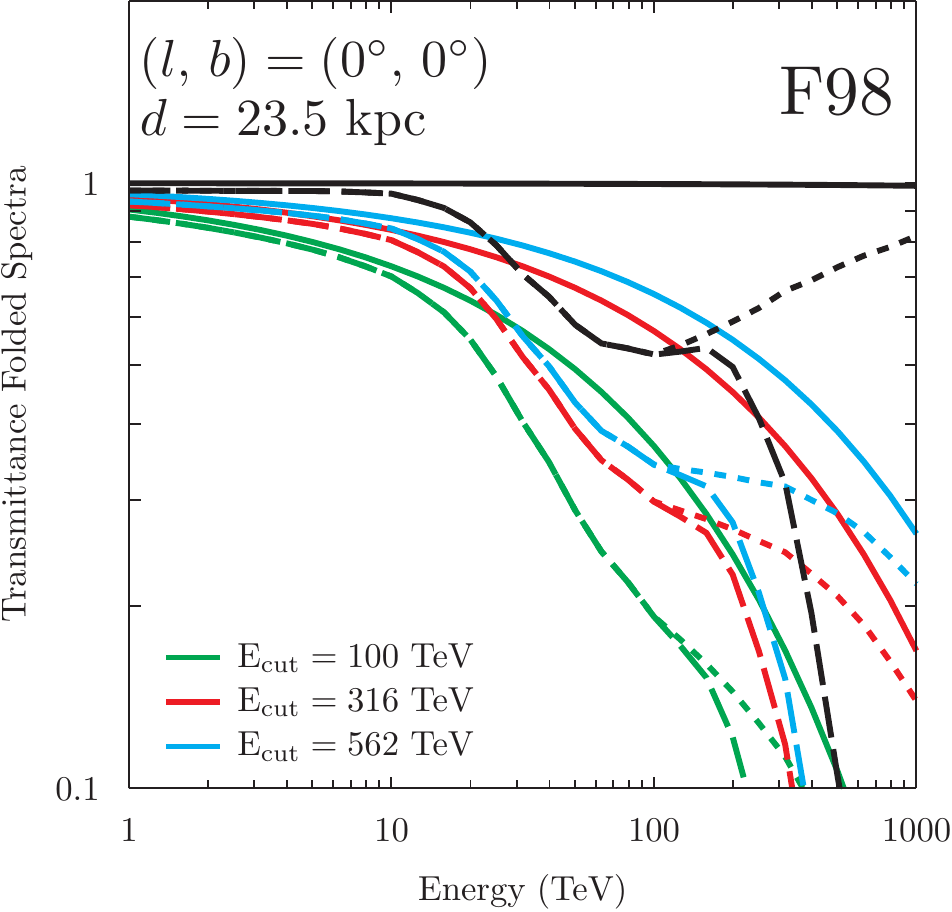}
  }
  \caption{Transmittance folded spectra for the R12 (left) and F98 (right) ISRF models for a source located at a distance $8.5$~kpc (top) and $23.5$~kpc (bottom) for the direction $(l,b) = (0^\circ, 0^\circ)$ in Galactic coordinates for selected exponential cutoff energies.
    Line styles: solid, no attenuation; short-dashed, ISRF-only attenuation; long-dashed, ISRF+CMB attenuation.
    Colours other than black are for cutoff energies $E_{\rm cut}$: green, 100~TeV; cyan, 316~TeV; red, 562~TeV.
    Note that the solid curves for all panels are identical.
  \label{fig:taueff}}
\end{figure*}

The spectral intensity for either ISRF model at each point in space is represented using a HEALPix Nside = 8 map with 256 logarithmic wavelength grid points per pixel from 0.1 to 10000~$\mu$m.
The spatial sampling is over a Galactocentric cylindrical grid with spacing that is variable in $R$ and $Z$, and remains constant azimuthally. (There are 44 radial bins sampling 0 to 30~kpc, 1 Z-bin for the plane and 8 additional above/below sampling to $\pm20$~kpc, and 36 azimuthal bins.)
Evaluation of the spectral intensity at all points $x$ in Eq.~(\ref{eq:tau}) for this component is made for either model using tri-linear interpolation.
The CMB is modelled as a spatially constant blackbody with temperature $T_{\rm CMB} = 2.725$~K.

Figure~\ref{fig:tau} shows the optical depth for the ISRF models at a \gray{} energy of 100~TeV for an integration path length of 23.5~kpc, the far side of the Galaxy stellar disc.
This is within the expected distance that the forthcoming CTA facility is expected to be able to discriminate spectral features $\gtrsim 10$~TeV \citep{2017arXiv170907997C}.
For this distance the optical depth is non-negligible for Galactic longitudes $-90^\circ \lesssim l \lesssim 90^\circ$ and latitudes $-45^\circ \lesssim b \lesssim 45^\circ$, and its distribution on the sky reflects that of the infrared photons predicted by the two ISRF models inside the Solar circle.
The optical depth is highest for the R12 model for longitudes $l\sim \pm 30^\circ$, which are lines of sight intersecting where the starlight from the spiral arms combines with the peak of the dust density distribution to produce the maximum of infrared emissions around $R \sim 4$~kpc, as discussed above.
Similarly, the higher infrared emissions for the F98 model over the inner Galaxy produce the correspondingly larger optical depths that peak toward the GC.
For both ISRF models there is a small amount of asymmetry in the optical depth maps about the $l = 0^\circ$ meridian caused by the spiral arms (R12) or stellar bar (F98) -- see Fig.~7 from \citet{2017ApJ...846...67P} for the spatial distribution of their integrated energy densities at the Galactic plane, which illustrates the asymmetrical features for the two ISRF models.

Figure~\ref{fig:taueff} shows the transmittance ($\exp[-\tau_{\gamma\gamma}(E)]$) for both ISRF models toward the GC at two distances folded with a sub-exponential cutoff function $\propto \exp(-[E/E_{\rm cut}]^{0.5})$ following \citet{2006PhRvD..74c4018K} with $E_{\rm cut} = 100$, 316, 562~TeV, and the no-cutoff case, respectively.
Here the underlying CR proton power-law spectrum has an exponential term $\propto \exp(-E/E_{\rm cut,p})$ with cutoff energy $E_{\rm cut,p}$ that is approximately an order of magnitude higher than $E_{\rm cut}$.
The unattenuated curves are shown as solid lines, with the broken lines showing the effect of ISRF-only (short-dashed) and combined ISRF/CMB (long-dashed) attenuation.
For a source located at the GC (using the IAU-recommended Sun-GC distance of 8.5~kpc \citep{1986MNRAS.221.1023K}) the F98 model produces about twice the attenuation compared to the R12 model around 100~TeV (transmittance $\sim$~0.85/0.7 for R12/F98).

The attenuation curves without cutoff illustrate the range of the likely effect for the ISRF models at different distances.
For a source located at the GC the pair-absorption effect mimics to some degree a spectrum with an intrinsic cutoff.
Only using information from the $10-100$~TeV energy range the \gray{} ``cutoff'' energy is $\sim500/1000$~TeV (F98/R12). 
For the case where the spectrum at source does have a cutoff the pair absorption steepens the spectrum further so that its observed shape appears as if the intrinsic cutoff has a lower \gray{} energy.
For a source located at the GC the downward shift for the inferred cutoff energy is between a factor $\sim 2$ (e.g., cyan and red long-dashed curves for R12) and $\sim 5$ (e.g., cyan and red long-dashed curves for F98).
For more distant sources the steepening can be more severe (see lower panels of Fig.~\ref{fig:taueff}).
Because \gray{} data has finite energy resolution extracting unique cutoff energies is non-trivial due to the different low-energy photon distribution for the ISRF models.

\begin{figure}[ht!]
  \includegraphics[scale=0.85]{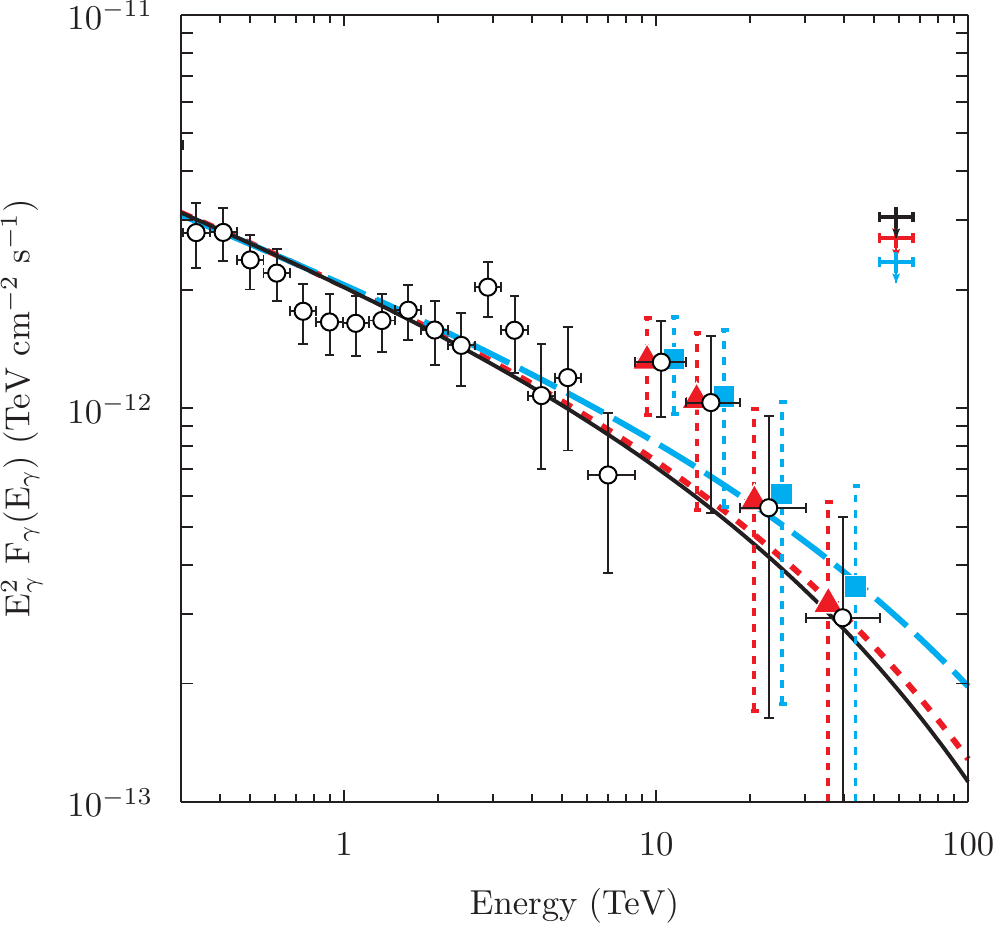}
  \caption{Spectrum of the ``diffuse'' emissions toward the GC measured by the HESS instrument \citep{2016Natur.531..476H} together with absorption-corrected data (note only absorption-corrected points $>10$~TeV are shown and these are offset compared to the original data by -10\% (R12) and +10\% (F98) in energy, respectively, for clarity).
    Point styles and colours: black open, uncorrected; red solid triangle, R12; cyan solid square, F98.
    Lines show the \gray{} spectral fit models used to estimate the 95\% lower confidence level to the proton cutoff energies (see text).
    Line types: solid, no ISRF correction; short-dashed, R12; long-dashed, F98.
  \label{fig:hessgc}}
\end{figure}

{\it Discussion:} Correction for the pair absorption always produces harder 
instrinsic spectra $\gtrsim 10$~TeV than observed.
Accounting for this effect can therefore affect the interpretation for the 
origin of \gray{} emissions from a source.
To see this the HESS data toward the GC for the
``diffuse'' spectrum attributed to the PeVatron there are ISRF-corrected and refit to obtain revised intrinsic spectral cutoff energy estimates.
Figure~\ref{fig:hessgc} shows the measured and ISRF-corrected data (note that the ISRF-corrected data are offset in energy by -10\% (R12) and +10\% (F98), respectively).
The {\tt naima} package \citep{naima2015} is used to derive the one-sided 68\% and 95\% lower confidence bands assuming a sub-exponential cutoff function for the \gray{} spectrum (the fitted power-law index for all three cases is found to be $\Gamma$=2.3)\footnote{The reduced $\chi^2$/dof $\sim$ 1 for a pure power-law fit. When a power-law + sub-exponential cutoff function is fitted a 
similar $\chi^2$/dof 
is also obtained, but with a cutoff beyond 1000~PeV (limited by the imposed cutoff parameter upper bound), hence the motivation 
to provide a confidence band on the cutoff lower limit as per the method 
outlined by \citet{2016Natur.531..476H}. The quality of the spectral fits is
not lowered when the pair-absorption correction is applied.}. 
Then, based on the procedure from 
\citet{2016Natur.531..476H}, {\tt naima} is also used to provide the underlying CR proton spectral cutoff energy 
for a \gray{} spectral model with $\Gamma=2.3$ and lowest cutoff energy fitting within each confidence band.
Figure~\ref{fig:hessgc} shows the \gray{} spectral models conforming to the 95\% confidence lowest cutoffs derived from the original and absorption-corrected data.
The resulting lower limits to the CR proton cutoff energies $E_{\rm cut,p}$ (TeV) are found to be (68\%/95\% = 2680/590; 3870/670; 5550/1180) for the no-ISRF, R12, and F98 cases, respectively.
The results obtained for the fits without ISRF model correction are consistent with the published values by \citet{2016Natur.531..476H} within the $\sim$15 to 20\% systematic error found from altering the {\tt naima} fit contraints and the method used to find the model function within each confidence band.
Thus it is found in this paper that the corrected $E_{\rm cut,p}$ limits are shifted higher by factors $\sim$1.3~to~2.1 (R12/F98), and the 95\% confidence lower limit for $E_{\rm cut,p}$ reaches beyond 1\,PeV for the F98 case.
Therefore, even though the pair-absorption correction provides only a modest upshift for the fluxes at the highest \gray{} energies measured for this source, the impact on the derived cutoff limit is non-negligible. 
If the CRs are linked to the central supermassive black hole, the increased CR proton cutoff energy will have follow-on implications for the parameters of the accelerator (e.g., magnetic field, black hole mass and/or acceleration region distance from the black hole -- see \citet{2005ApJ...619..306A}).
Such implications will become more apparent as future \gray{} observations probe deeply beyond 100~TeV energies.
Note that the hardening of the intrinsic spectrum $\gtrsim10$~TeV following 
the pair-absorption correction strengthens the case against a leptonic origin for the 
emissions, because the rapid cooling on the intense radiation and magnetic fields about the GC region produce softer spectra for this scenario.

Comparing the pair-absorption calculations with other recent works, the transmittance for a source located at the GC for the F98 model is comparable to that obtained by \citet{2017MNRAS.470.2539P} (their Fig.~15, left panel), while the R12 model is slightly lower than obtained by \citet{2016PhRvD..94f3009V} (their Fig.~12).
It should be emphasised, though, that R12 and F98 are equivalent solutions for the Galaxy-wide ISRF using fully 3D calculations with the same radiation transfer code but different stellar/dust density distribution, while achieving similar agreement with the near- to far-infrared observations.
As discussed earlier the models cannot be arbitrarily modified to produce much lower or higher infrared photon densities, particularly over the inner Galaxy, and the calculations made in this paper can therefore be considered to provide likely bounds on the pair attenuation relevant for TeV \gray{} measurements.

The forthcoming CTA TeV \gray{} facility is expected to detect and measure
spectra from sources right across the Galaxy
(see Sec.~10.4 of \citep{2017arXiv170907997C}).
Even for the ISRF model with the lowest absorption (R12) the correction 
will be important for assessing intrinsic spectral characteristics with \gray{} data up to $\sim 200$~TeV, below the energies where CMB attenuation is important, but where the ISRF attenuation factor can reach $\sim50$\% for a source on the other side of the Galaxy.
For the currently operating HAWC instrument with improvements in its
spectral reconstruction $\gtrsim 10$~TeV \citep{2017arXiv170803502S} the pair-absorption corrections may also be similarly important, given the expectation to detect sources beyond 100~TeV energies.
To aid assessments of the effect on VHE \gray{} spectra the full set of all-sky optical depth maps in energy and distance for the R12 and F98 models calculated in this paper will 
be available from the \GP\ website, https://galprop.stanford.edu.

Putting these points together, $\gtrsim 10$~TeV \gray{} observations with future
instruments of a population (100s as could be expected) of sources across the
Galaxy will also reveal the 3D structure of the ISRF.
Such observations can be used for optimising the description of the low-energy photon distribution in the Galaxy, providing complementarity to other studies that more commonly employ near- to far-infrared data.

{\it Acknowledgements:}
TAP and IM acknowledge partial support via NASA grant NNX17AB48G.

The {\tt naima} package is available at http://naima.readthedocs.io/en/latest/radiative.html.

Some of the results in this paper have been derived using the HEALPix \cite{2005ApJ...622..759G} package.


\bibliography{isrf.bib}

\end{document}